\documentclass{article}
\usepackage{spconf,amsmath,amssymb,graphicx,booktabs,flafter,fix2col,enumitem}
\usepackage[hidelinks]{hyperref}
\usepackage[table]{xcolor}

\title{Point-and-Play: Instrument Timbre Cloning from Music Mixtures}
\name{Baotong Tian, Zhiyao Duan}
\address{Department of Electrical and Computer Engineering, University of Rochester, Rochester, NY, USA}
\begin{document}
\ninept
\maketitle
\begin{abstract}
Neural instrument synthesizers can clone the timbre of a solo instrument, letting music producers make music with the sound they desire. However, an instrument of interest is often embedded in a fully produced, multi-instrument recording, beyond the reach of existing methods. Source separation before timbre cloning is an approach, but it may suffer from separation artifacts and is also likely a detour for the task. In this work, we propose \textbf{Point-and-Play}, a system that clones the timbre of a cued source directly from a polyphonic mixture and uses it to synthesize new pitches and velocities. 
To achieve this, we introduce a Cue-Gated Statistics Pooling (CGSP) module that extracts the timbre code of the cued instrument from adapted CLAP embeddings of the music mixture. This extracted timbre code is then used to condition a separately trained note synthesizer together with pitch and velocity inputs. 
Experiments show that our proposed system outperforms separate-then-clone baselines in terms of timbre fidelity and synthesis quality, achieving comparable performance with that cloned from a clean solo instrument\footnote{Code and demo page are available at \href{https://github.com/User-tian/Point-and-Play}{\url{https://github.com/User-tian/Point-and-Play}}.}.

\end{abstract}
\section{Introduction}
\label{sec:intro}
Neural instrument synthesis can render realistic notes from pitch and velocity controls, with instrument timbre specified by a text description or an isolated audio reference~\cite{nercessian2024,tokensynth,flowsynth}. Instrument-specific systems have likewise improved the fidelity and expressive control of score- and MIDI-driven synthesis for piano~\cite{performancenet,diffpiano,midivalle}, guitar~\cite{neuralguitar,guitarflow}, and violin~\cite{violindiff,violet}. A musician, however, may want to clone a particular instrument heard only within a polyphonic recording and use its timbre in music production. 
Existing approaches can isolate the desired stem through source separation~\cite{demucs,clapsep,promptsep}, or replace the timbre of a source in a music mixture with a reference timbre~\cite{dismix}, yet neither formulation addresses cloning the timbre of an instrument source from the mixture to synthesize new music content.

\begin{figure}[t]
\centering
\includegraphics[width=\columnwidth]{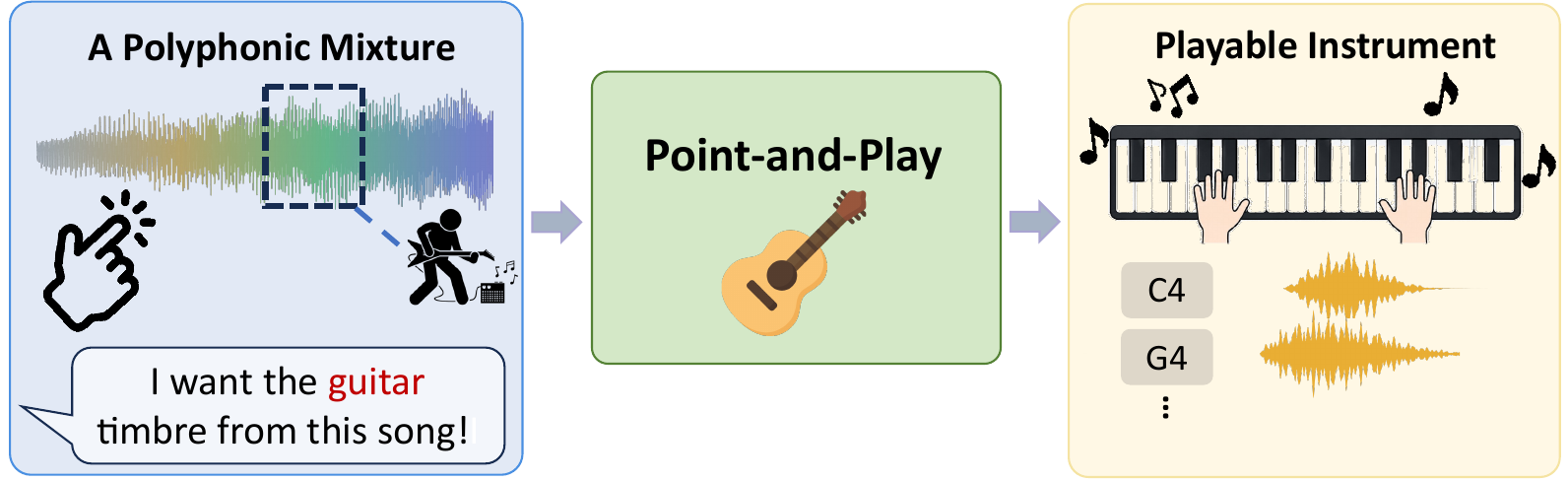}
\caption{Overview of Point-and-Play. Given a polyphonic mixture and an audio or text cue, the system extracts the cued source's timbre  from the mixture and uses it to render new pitches and velocities.}
  \label{fig:overview}
\vspace{-2em}
\end{figure}

Recovering a reusable timbre representation from such material is challenging because information about the target instrument is entangled with its pitch, recording conditions, and competing sources. A natural approach is to first separate the target source from the mixture and then apply a timbre cloning model designed for isolated references. However, many source separation systems are designed to recover a predefined set of stems~\cite{waveunet,demucs,melbandroformer}, making them ill-suited to arbitrary instruments specified by a cue. Even when a suitable separator is available, separation artifacts can propagate to the downstream synthesizer and degrade timbre fidelity. Moreover, waveform reconstruction is an unnecessary intermediate objective for this use case, since the separated stem itself is not the desired output. 
We therefore pursue direct mixture encoding, in which the target timbre is inferred directly from the mixture without the intermediate target reconstruction. The main challenge is to make this representation selective to the cued source while discarding both competing sources and source-specific musical content such as pitch, so that the resulting timbre can be reused to synthesize new notes.

We formalize this missing task as \emph{instrument cloning from music mixtures}. Given a multi-instrument polyphonic mixture, the task is to clone the timbre of an instrument source cued by an audio example of the target instrument class or a text prompt containing its instrument name, and use it to synthesize music content that does not appear in the mixture. 
This distinguishes the task from timbre cloning from an isolated instrument recording, where the target timbre is directly observed without interference from other sources, and from text-to-instrument generation, where text specifies the desired instrument rather than identifying a source whose timbre must be recovered from a mixture.

To address this task, we introduce \textbf{Point-and-Play}, which extracts a cued instrument's timbre directly from a mixture. Specifically, we propose Cue-Gated Statistics Pooling (CGSP), which uses the cue to aggregate adapted CLAP~\cite{clap} audio representations of the mixture into a timbre code. We train the extracted code to align with the CLAP embedding of the corresponding instrument in isolation, while encouraging invariance to musical content and selectivity to the cue. Our contributions are threefold:
\begin{enumerate}[
label=\arabic*., leftmargin=1.7em, labelsep=0.5em, itemsep=0pt, parsep=0pt, topsep=2pt ]
\item We formulate a new task, \emph{instrument cloning from music mixtures}, where a cued instrument is cloned from a mixture, and establish an evaluation protocol for this task.

\item We introduce \textbf{Point-and-Play}, the first system, to our knowledge, to clone a cued source from a polyphonic mixture toward synthesizing new pitches and velocities.

\item Experiments show that 
Point-and-Play achieves timbre fidelity comparable to instrument cloning from a clean audio reference while outperforming a separate-then-clone baseline.
\end{enumerate}

\section{Related Work}
\label{sec:related}
Text-to-instrument generation and instrument cloning aim to create reusable timbres that can be rendered at new pitches and velocities. Earlier systems derive instrument timbres from isolated examples~\cite{ganstrument,fewsampleddsp}, while recent work has broadened timbre conditioning and improved the generative models. InstrumentGen generates sample-based instruments from text by combining a joint text--audio embedding with pitch and velocity controls~\cite{instrumentgen,nercessian2024}. TokenSynth achieves this task by generating neural audio tokens from MIDI and CLAP-based timbre embeddings~\cite{tokensynth}. More recently, FlowSynth improves single-note quality and cross-pitch timbre consistency through distributional flow matching and test-time search~\cite{flowsynth}. Our system shares the goal of producing a reusable, playable timbre, but extracts it from a cued source embedded in a polyphonic mixture.

Music source separation instead reconstructs sources present in a mixture. Conventional systems target a fixed vocabulary, typically vocals, drums, bass, and ``other''~\cite{waveunet,openunmix,demucs}, a setup largely retained by newer architectures~\cite{lasaft,bandsplitrnn,melbandroformer}. Target sound extraction enables user-selected sources via class labels or enrollment audio~\cite{soundbeam}, language~\cite{lassnet,clipsep,audiosep}, and multimodal queries~\cite{clapsep}. Recently, PromptSep supports open-vocabulary extraction and removal via language or vocal imitation~\cite{promptsep}. For our task, one could first separate the target instrument and then clone its timbre, but source separation methods face a tradeoff between separation quality and source diversity.

\section{Method}
\label{sec:method}

\subsection{Cue-Conditioned Timbre Extraction}

Given a polyphonic music mixture $x_{\mathrm{mix}}$ and a source cue, Point-and-Play extracts a timbre code $z$ for the cued instrument directly from the mixture. The cue can be an audio reference or a text cue, and the resulting timbre code is later used together with the requested pitch and velocity to condition the note synthesizer.

\begin{figure*}[t]
\centering
\includegraphics[width=0.9\textwidth]{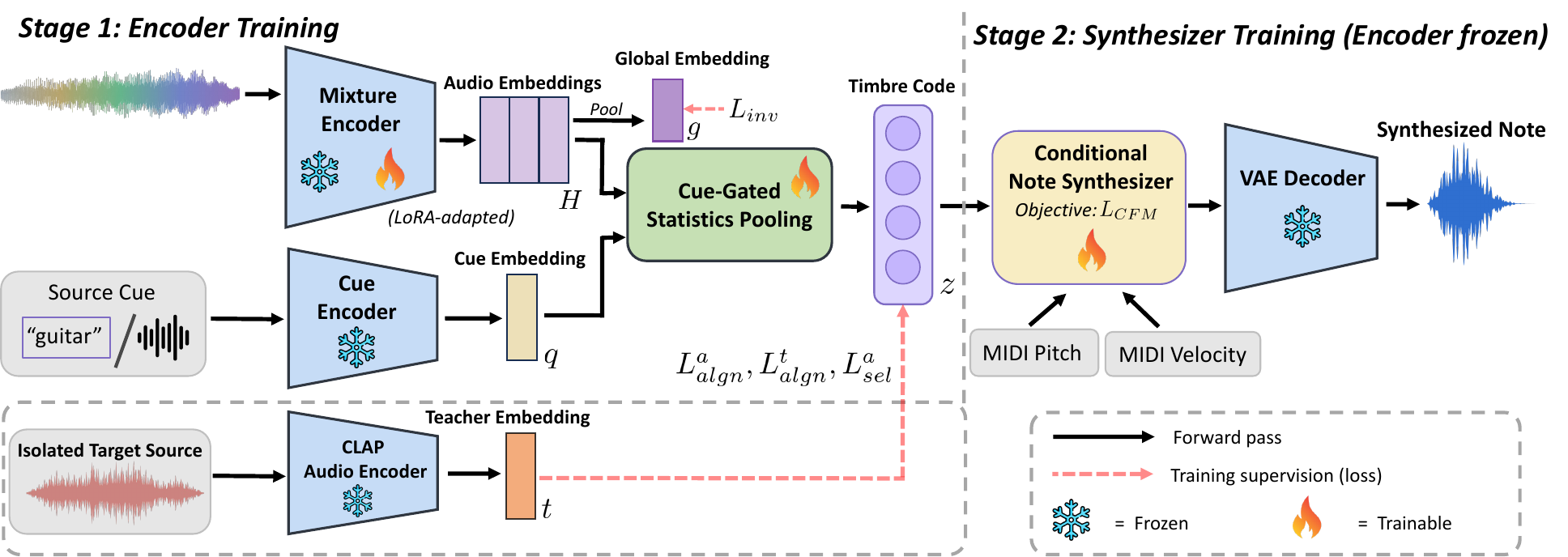}
\caption{Point-and-Play architecture. A CLAP encoder fine-tuned with LoRA encodes the mixture, while a frozen CLAP encoder encodes the source cue. Cue-Gated Statistics Pooling combines both representations into a timbre code of the cued source in the mixture and uses it to condition the note synthesizer. During training, the isolated source of the mixture provides supervision; it is not required at inference time.}
  \label{fig:architecture}
\vspace{-2mm}
\end{figure*}


As shown in Fig.~\ref{fig:architecture}, the timbre extraction module consists of two encoding branches followed by a Cue-Gated Statistics Pooling (CGSP) module. The trainable mixture branch uses a CLAP audio encoder fine-tuned with Low-Rank Adaptation (LoRA)~\cite{clap,lora}, which represents the mixture by a global code $g$ and token sequence $H=[h_1,\ldots,h_T]$. In parallel, the frozen cue branch uses a CLAP audio/text encoder to map the audio or text cue into the shared audio--text embedding space, producing a cue embedding $q$. CGSP then uses $q$ to selectively aggregate the mixture tokens $H$ to produce the final timbre code $z$. 

Specifically, CGSP produces $z$ in three stages: cue--token gating, weighted statistics aggregation, and residual fusion. The first stage projects each mixture token $h_i$ and the cue $q$ into a shared $d$-dimensional space with $W_h$ and \(W_q\), respectively, and computes a cue-conditioned gate $a_i\in[0,1]$ for each token using a sigmoid function. Unlike softmax attention weights, these gates are computed independently, allowing multiple time indices to receive large weights. The second stage aggregates the mixture tokens using the $a_i$-weighted mean $\mu$ and standard deviation $\sigma_h$. Finally, the multilayer perceptron (MLP) $R_\theta$ maps $[\mu;\sigma_h]$ to a mixture-conditioned residual, which is added to the $\ell_2$-normalized cue embedding. The resulting vector is again $\ell_2$-normalized to obtain the timbre code $z$:
\begin{equation}
\begin{aligned}
a_i&=\operatorname{sigmoid}\!\left((W_hh_i)^\top W_qq/\sqrt d\right),\\
\mu&=\frac{\sum_i a_i h_i}{\sum_i a_i},\qquad
  \sigma_h=\sqrt{\frac{\sum_i a_i(h_i-\mu)^2}{\sum_i a_i}},\\
z&=\operatorname{norm}\!\left(\operatorname{norm}(q)+R_\theta[\mu;\sigma_h]\right).
\end{aligned}
\label{eq:cgsp}
\end{equation}
We zero-initialize the final layer of $R_\theta$, so the extracted timbre code $z$ initially represents the cue prior and gradually learns the mixture evidence that distinguishes the selected instrument.

\subsection{Extractor Training Objectives}
\label{ssec:losses}
For each target source $s$, we use a frozen CLAP audio encoder to obtain
a teacher embedding $t=C_a(s)$. Given the corresponding mixture, the timbre
extractor produces an audio-cued timbre code $z^a$ and a text-cued code $z^t$ following Eq.~\eqref{eq:cgsp}.
We train the extractor with four complementary losses and use standard symmetric InfoNCE~\cite{oord2018,clap} with temperature $\tau$:
\begin{align}
\mathcal L_{\rm inv}
    &= \mathrm{InfoNCE}_{\tau}(g^A,g^B), &
\mathcal L_{\rm algn}^{a}
    &= \mathrm{InfoNCE}_{\tau}(z^a,t), \nonumber\\
\mathcal L_{\rm algn}^{t}
    &= \mathrm{InfoNCE}_{\tau}(z^t,t), &
\mathcal L_{\rm sel}^{a}
    &= \mathbb E
        [\,S(\bar z^a,t)-m\,]_+ .
\label{eq:component-losses}
\end{align}
For $\mathcal L_{\rm inv}$, $g^A$ and $g^B$ are global embeddings of
paired mixtures containing the same instruments but different
musical content. Aligning them encourages the mixture encoder to be invariant to the musical content.
The alignment losses $\mathcal L_{\rm algn}^{a}$ and
$\mathcal L_{\rm algn}^{t}$ instead align the cue-conditioned codes, $z^a$ and $z^t$, with
the isolated target embedding $t$, respectively. For the audio condition, the cue is drawn
from a different instrument in the same category, requiring the model to
extract the instrument timbre from the mixture rather than copy the cue
timbre.
To encourage cue selectivity, $\mathcal L_{\rm sel}^{a}$ uses a
mismatched audio cue (i.e., audio cue of a different instrument category than the target) to extract a timbre code $\bar z^a$ from the same mixture, and pushes it away from the target embedding $t$. Here $S$ denotes cosine similarity and $[\cdot]_+$ is the hinge operator. 
We do not apply this loss to text cues because they are categorical in this work; hence, pushing negative cue embeddings away is not likely to provide additional benefits over $\mathcal L_{\rm algn}^{t}$. 
The full extractor objective is a weighted sum
\begin{equation}
\mathcal L_E =
\lambda_{\rm inv}\mathcal L_{\rm inv}
+\lambda_a\mathcal L_{\rm algn}^{a}
+\lambda_t\mathcal L_{\rm algn}^{t}
+\lambda_s\mathcal L_{\rm sel}^{a},
\label{eq:loss}
\end{equation}
where the $\lambda$ coefficients control the contribution of each term.
Table~\ref{tab:ablation} ablates the four objectives.

\subsection{Conditional Note Synthesizer and Inference}
\label{ssec:synth}

To convert the timbre code into a playable instrument, we freeze the extractor $E_\theta$ and train a latent note synthesizer inspired by FlowSynth~\cite{flowsynth}. The synthesizer is conditioned jointly on the extracted timbre code $z$, pitch $p$, and velocity $v$, through $c=W_z z+E_p(p)+E_v(v)$. Here, $W_z$ is a learned linear projection of the 512-dimensional timbre code, while $E_p(\cdot)$ and $E_v(\cdot)$ are learned embedding functions for pitch and velocity, respectively. Their element-wise sum $c$ modulates each DiT block through Adaptive Layer Normalization (AdaLN)~\cite{dit}. Given a target audio latent $y$, Gaussian noise $\epsilon$, and flow time $u\sim\mathcal U[0,1]$, we form the linear path $y_u=(1-u)\epsilon+uy$ and optimize the standard Conditional Flow-Matching (CFM) loss
\begin{equation}
\mathcal L_{\rm CFM}=\mathbb E_{(y,c),u,\epsilon}\lVert F_\phi(y_u,u,c)-(y-\epsilon)\rVert_2^2,
\label{eq:flow}
\end{equation}
where $F_\phi$ is the Transformer trained to predict the path velocity $y-\epsilon$ from the intermediate latent $y_u$, flow time $u$, and condition $c$. At inference, the same condition specifies the selected instrument with a new pitch and velocity. We then integrate $F_\phi$ from Gaussian noise to obtain the latent and use the frozen VAE decoder to decode the waveform. We generate multiple candidates and keep the one whose frozen CLAP embedding has the highest cosine similarity to $z$.

\section{Experiments}
\label{sec:exp}


\subsection{Experimental Setup}
\label{ssec:setup}

\begin{figure}[t]
\centering
\includegraphics[width=0.7\columnwidth]{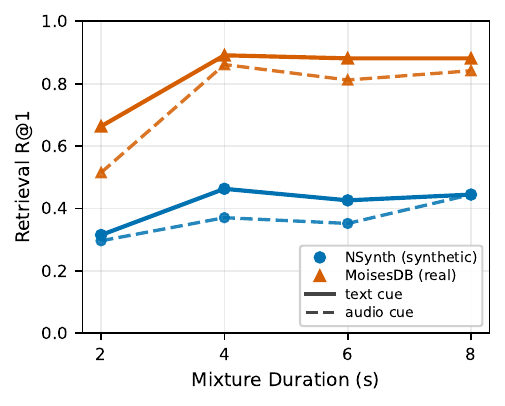}
\vspace{-5mm}
\caption{Decoder-free paired retrieval Recall@1 versus mixture duration on NSynth and MoisesDB. Only the centered crop duration varies. Solid and dashed lines denote text and same-category audio cues, respectively; colors and markers distinguish datasets.}
  \label{fig:duration}
\vspace{-5mm}
\end{figure}

\noindent\textbf{Data.} The extractor is trained on polyphonic mixtures drawn from four corpora, namely NSynth~\cite{nsynth}, NLakh~\cite{showmeinstruments}, Slakh~\cite{slakh}, and MoisesDB~\cite{moisesdb}, sampled with weights $0.30/0.30/0.10/0.30$. For NSynth and NLakh, we randomly select four of 953 instruments and independently draw two sets of 4-second excerpts to form paired mixtures that share instrument identities but differ in musical content.
For Slakh and MoisesDB, the paired mixtures $x_{\rm mix}^A$ and $x_{\rm mix}^B$ comprise two independently sampled 4-second windows from the same song, retaining all available stems. Thus, all mixtures used for extractor training are 4 seconds long. All evaluations in Table~\ref{tab:main}, Figure~\ref{fig:duration}, and Table~\ref{tab:ablation} use 27 NSynth instruments held out from training. Figure~\ref{fig:duration} additionally uses an evaluation subset of MoisesDB excluded from extractor training.
\begin{table}[t]
\centering
\caption{Generation results on 27 held-out NSynth identities. CLAP score is the generated-to-target audio similarity. Arrows show the preferred direction, and bold marks the best Point-and-Play value.}
\label{tab:main}
\footnotesize
\setlength{\tabcolsep}{3pt}
\begin{tabular}{@{}lccc@{}}
\toprule
System and conditioning input & CLAP $\uparrow$ & FAD $\downarrow$ & MSS $\downarrow$ \\
\midrule
\rowcolor[gray]{0.92}\multicolumn{4}{@{}l}{\emph{Oracle Timbre}}\\
Target note       & 0.811 & 1.78 & 2.04 \\
Target-instrument other note & 0.762 & 1.85 & 2.38 \\
\midrule
\rowcolor[gray]{0.92}\multicolumn{4}{@{}l}{\emph{Cue Timbre}}\\
Text cue                  & 0.548 & 2.27 & 3.84 \\
Target-category audio cue     & 0.584 & 1.82 & 3.25 \\
Target-category centroid    & 0.634 & 2.62 & 3.17 \\
\midrule
\rowcolor[gray]{0.92}\multicolumn{4}{@{}l}{\emph{Separate-then-clone}}\\
Oracle ratio mask, GT stem       & 0.726 & 2.20 & 2.51 \\
Demucs-closest~\cite{demucs}, fixed stem & 0.541 & 4.13 & 3.55 \\
CLAPSep~\cite{clapsep}, query    & 0.560 & 2.38 & 3.55 \\
\midrule
\rowcolor[gray]{0.92}\multicolumn{4}{@{}l}{\emph{Point-and-Play}}\\
Text cue                         & \textbf{0.706} & 2.23          & 2.68 \\
Same-category audio cue          & 0.678          & \textbf{1.92}          & 2.75 \\
Target-instrument audio cue      & 0.696          & \textbf{1.92} & \textbf{2.64} \\

\bottomrule
\end{tabular}
\vspace{-3mm}
\end{table}

\noindent\textbf{Implementation.} The mixture encoder is initialized from the HTSAT-base audio tower of a music-domain LAION-CLAP checkpoint~\cite{clap,laionclapcheckpoints} and fine-tuned with rank-8 LoRA~\cite{lora} ($\alpha{=}16$) in every qkv, fc1, and fc2 projection; only the LoRA parameters and CGSP head are trained. The frozen cue and teacher encoders, as well as the candidate-selection embeddings, use the same music-specific checkpoint. For text-cued training and evaluation, each text cue is a coarse instrument-category name, such as ``guitar''. All mixture views are cropped to four seconds during extractor training. We train the extractor for 18k steps with AdamW, a batch size of 16, and an InfoNCE temperature of $0.1$. The learning rate is warmed up for 500 steps to $3{\times}10^{-4}$ and cosine-decayed to $3{\times}10^{-5}$. Training takes approximately three hours on a single A100 GPU. The loss weights are $\lambda_{\rm inv}{=}\lambda_a{=}\lambda_t{=}1$ and $\lambda_s{=}0.25$. The note synthesizer follows the FlowSynth latent flow-matching decoder~\cite{flowsynth}, using a 24-layer DiT with a hidden width of 1,024, 16 attention heads, and AdaLN conditioning. The 512-dimensional timbre code is linearly projected to the 1,024-dimensional DiT width and summed with pitch and velocity embeddings to form the conditioning vector. The model predicts continuous latents in the DiffRhythm VAE space~\cite{diffrhythm}, which are converted to waveforms by the frozen VAE decoder. It is trained with the flow-matching objective in Eq.~\eqref{eq:flow} for 150k steps using AdamW, a learning rate of $10^{-4}$, batch size 48, and condition dropout of $0.1$ for classifier-free guidance. Conditioning codes are extracted by the final encoder and sampled equally from NSynth note mixtures and NLakh phrases. At inference, we integrate 50 Euler steps from Gaussian noise and, for each conditioning code $z$, generate $N{=}8$ samples and retain the one whose frozen music-CLAP embedding has the highest cosine similarity to $z$.

\noindent\textbf{Comparisons.} Table~\ref{tab:main} compares four groups of settings or system configurations.
\textbf{Oracle timbre.} These settings condition the synthesizer with a Music-CLAP embedding extracted from clean recordings of the target instrument, setting oracle upper bounds. Specifically, \emph{target note} uses the exact synthesis target note to compute the embedding, 
while \emph{target-instrument other note} uses a different note from the target instrument. 
\textbf{Cue timbre.} These settings omit the mixture and simply use cue embeddings to condition the synthesizer. Specifically, \emph{text cue} uses the name of the target instrument category, 
\emph{target-category audio cue} uses a recording of a different instrument from the target instrument category, and \emph{target-category centroid} uses the average embedding of all training recordings in the target instrument category. These settings serve as baselines to test whether the cue (or average of cues) alone is sufficient to infer the target instrument timbre.
\textbf{Separate-then-clone.} These systems first separate the target source from the mixture and then condition the synthesizer using the Music-CLAP embedding of the separated source. These systems differ by their separation methods. 
\emph{Oracle ratio mask} uses a mask computed from the ground-truth stem to separate the source. 
\emph{Demucs-closest} separates the mixture into four stems using a pre-trained Demucs model~\cite{demucs} and uses the one whose CLAP embedding is the closest to the target instrument as the separation result. 
\emph{CLAPSep}~\cite{clapsep} embeds a text or audio cue in CLAP's shared embedding space and uses it to condition a separation network that estimates a mask for extracting the queried source.
\textbf{Point-and-Play.} We evaluate three cue types for extracting a timbre code directly from the mixture: a \emph{text cue} specifying the target instrument category name, a \emph{same-category audio cue} from a different instrument in the same category, and a \emph{target-instrument audio cue} from the exact target instrument.
In particular, the first two cues are most practical. For all systems, the synthesizer is trained to match the timbre space that the systems use. 
Because the conditioning spaces differ, the oracle timbre, cue timbre, and separate-then-clone comparisons use a decoder trained on Music-CLAP embeddings of isolated sources, whereas Point-and-Play uses the same decoder architecture trained on timbre codes extracted from mixtures by CGSP.


\noindent\textbf{Metrics.} The primary metric is \emph{CLAP score}, computed as the cosine similarity between each synthesized note and its synthesis target using a separate HTSAT-tiny audio tower of a general-audio LAION-CLAP checkpoint~\cite{clap,laionclapcheckpoints}. 
\emph{FAD}~\cite{fad} measures distribution-level realism from VGGish embeddings at 16\,kHz, and \emph{MSS} is a multi-resolution spectral distance, with lower values preferred for both. 
\begin{table}[t]
\centering
\caption{Ablation of the encoder objective in Eq.~\eqref{eq:loss}. All reported values are CLAP scores. Txt and Aud use the matched category text and a clean same-category audio cue from a different instrument, respectively; W$_t$ and W$_a$ use mismatched text and audio cues. Lower wrong-cue scores indicate stronger cue selectivity.}
\label{tab:ablation}
\begin{tabular}{@{}lcccc@{}}
\toprule
Objective & Txt $\uparrow$ & Aud $\uparrow$ & W$_t\downarrow$ & W$_a\downarrow$ \\
\midrule
Full objective (ours)             & \textbf{0.699} & \textbf{0.708} & 0.418 & 0.430 \\
\quad $-\,\mathcal L_{\rm inv}$      & 0.654 & 0.640 & 0.399 & 0.428 \\
\quad $-\,\mathcal L_{\rm algn}^{a}$ & 0.661 & 0.417 & 0.399 & 0.400 \\
\quad $-\,\mathcal L_{\rm algn}^{t}$ & 0.578 & 0.646 & 0.425 & 0.403 \\
\quad $-\,\mathcal L_{\rm sel}^{a}$  & 0.664 & 0.664 & 0.424 & 0.469 \\
\bottomrule
\end{tabular}
\vspace{-3mm}
\end{table}

\subsection{Evaluation Results}

\noindent\textbf{Playable-instrument generation from mixtures.} 
Table~\ref{tab:main} reports the main generation results. Point-and-Play achieves a CLAP score of $0.706$ with text cue, $0.678$ with a same-category audio cue, and $0.696$ with target-instrument audio cue. These values are not far from the ``target instrument other note'' oracle of $0.762$, which shows that directly extracting timbre from the mixture with our proposed method can approach the performance of conditioning on a clean recording of the target instrument. Moreover, the two audio-cued configurations achieve an FAD of $1.92$, close to the clean-reference oracle of $1.85$, indicating strong distribution-level generation quality. Point-and-Play also outperforms all ``cue timbre'' methods in terms of CLAP score and MSS, showing the benefits of extracting the target-instrument-specific timbre from the mixture over using the timbre of a different instrument or the average timbre of the same category. Compared with separate-then-clone methods, all configurations of Point-and-Play outperform Demucs and CLAPSep across all metrics, suggesting that direct encoding preserves target timbre more effectively than the tested two-stage pipelines.

\noindent\textbf{Pitch leakage investigation.}
To assess whether the extracted representation retains the pitch observed in the mixture, we fit a five-fold cross-validated ridge regressor on held-out NSynth instruments to predict the MIDI pitch of the cued target note from the extracted timbre code, and report $R^2$ (higher indicates more linearly decodable pitch information). The same probe achieves $R^2=0.783$ on raw music-CLAP embeddings of the corresponding isolated target notes, compared with $R^2=0.263$ on the extracted codes, indicating that the extractor suppresses, but does not eliminate, pitch information.

\noindent\textbf{Mixture-duration study.} 
We evaluate mixture-duration sensitivity using decoder-free Recall@1, which measures whether the extracted code retrieves the correct clean target from a fixed pool of instruments. Figure~\ref{fig:duration} shows that two seconds is insufficient on both datasets, while four seconds captures most of the retrieval gain, and performance remains stable with longer context. For NSynth, each sample is a single four-second note, the result therefore indicates that one note is sufficient for retrieval. We therefore use four-second crops by default, consistent with the training window. 

\noindent\textbf{Loss ablation.} Table~\ref{tab:ablation} reports leave-one-out ablations of the full objective (Eq.~\eqref{eq:loss}). All variants are evaluated using a single fixed decoder, trained exclusively on audio-CLAP embeddings of clean instrument recordings, ensuring that performance differences primarily reflect the encoder. We can see that the audio selectivity loss $\mathcal L_{\rm sel}^{a}$ improves the matched-cue CLAP score for both cue modalities while reducing the wrong-audio-cue score W$_a$ from $0.469$ to $0.430$. It therefore enhances both target fidelity and selectivity to the audio cue. The instance-level losses are essential for preserving identity under their respective cue modalities. Specifically, removing $\mathcal L_{\rm algn}^{a}$ reduces audio-cued CLAP score to $0.417$, close to the wrong-cue score, whereas removing $\mathcal L_{\rm algn}^{t}$ reduces text-cued CLAP score to $0.578$. In both cases, performance under the other cue remains largely unaffected, indicating that each loss provides modality-specific supervision. Finally, removing $\mathcal L_{\rm inv}$ degrades performance under both cues. This result suggests that encouraging timbre representation invariance to pitch variation improves timbre capture from polyphonic mixtures across cue modalities. 

\section{Conclusion}
\label{sec:conc}
In this paper, we introduced a new task of \emph{instrument cloning from music mixtures} and presented \textbf{Point-and-Play}, a system that turns a selected source within a polyphonic mixture into a playable instrument at new pitches and velocities. By extracting a timbre representation directly from the mixture, Point-and-Play approaches the performance of clean-reference conditioning and outperforms separate-then-clone pipelines. Future work will improve real-world generalization by scaling training to more diverse instruments and recordings and further assess perceptual quality through listening studies. Another important direction is to enable finer-grained source selection, particularly when multiple simultaneous instruments belong to the same category.

\begingroup
\let\OLDthebibliography\thebibliography
\renewcommand\thebibliography[1]{%
  \OLDthebibliography{#1}%
  \setlength{\itemsep}{0pt}%
  \setlength{\parsep}{0pt}%
  \setlength{\topsep}{0pt}%
  \linespread{0.92}\selectfont%
}
\bibliographystyle{IEEEbib}
\bibliography{strings,refs}
\endgroup

\end{document}